\documentclass[conference]{IEEEtran}%
\usepackage{amsfonts}
\usepackage{amsmath}
\usepackage{amssymb}
\usepackage{graphicx}%
\IEEEoverridecommandlockouts
\providecommand{\U}[1]{\protect\rule{.1in}{.1in}}
\newtheorem{theorem}{Theorem}

\newtheorem{lemma}[theorem]{Lemma}

\newtheorem{remark}[theorem]{Remark}

\newenvironment{proof}[1][Proof]{\noindent\textbf{#1.} }{\ \rule{0.5em}{0.5em}}
\begin{document}

\title{Explicit capacity-achieving receivers for optical communication and quantum reading}
\author{\IEEEauthorblockN{Mark M. Wilde\IEEEauthorrefmark{1},
Saikat Guha\IEEEauthorrefmark{2},
Si-Hui Tan\IEEEauthorrefmark{3}, and 
Seth Lloyd\IEEEauthorrefmark{4}}
\IEEEauthorblockA{\IEEEauthorrefmark{1}School of Computer Science, McGill University, Montreal, Quebec H3A 2A7,
Canada}
\IEEEauthorblockA{\IEEEauthorrefmark{2}Disruptive Information Proc. Tech. Group, Raytheon BBN
Technologies, Cambridge, Massachusetts 02138, USA}
\IEEEauthorblockA{\IEEEauthorrefmark{3}Data Storage Institute, Agency for Science, Tech., \& Research, 117608 Singapore}
\IEEEauthorblockA{\IEEEauthorrefmark{4}Research Laboratory for Electronics and
Department of Mechanical Engineering,\\
Massachusetts Institute of Technology,
Cambridge, Massachusetts 02139, USA}}

\maketitle

\begin{abstract}
An important practical open question has been to design explicit, structured optical receivers that achieve the Holevo limit in the contexts of optical communication and ``quantum reading.'' The Holevo limit is an achievable rate that is higher than the Shannon limit of any known optical receiver. We demonstrate how a sequential decoding approach can achieve the Holevo limit for both of these settings. A crucial part of our scheme for both settings is a non-destructive \textquotedblleft vacuum-or-not\textquotedblright\ measurement that projects an $n$-symbol modulated codeword onto the $n$-fold vacuum state or its orthogonal complement, such that the post-measurement state is either the $n$-fold vacuum or has the vacuum removed from the support of the $n$ symbols' joint quantum state. The sequential decoder for optical communication requires the additional ability to perform multimode optical phase-space displacements---realizable using a beamsplitter and a laser, while the sequential decoder for quantum reading also requires the ability to perform phase-shifting (realizable using a phase plate) and online squeezing (a phase-sensitive amplifier).

\end{abstract}

One of the first accomplishments in quantum information theory was the upper
bound (now known as the {\em Holevo bound}) on how much classical information can
be encoded into a quantum system, such that another party can reliably recover
it using a quantum measurement \cite{Holevo73}. Subsequently, Holevo, Schumacher,
and Westmoreland (HSW) proved that the Holevo bound is also an achievable
rate for classical communication over a quantum channel~\cite{ieee1998holevo,SW97}, establishing a lower bound on a quantum channel's classical capacity. These initial results were the impetus for the field of quantum information theory~\cite{W11}, a generalization of Shannon's classical information theory that takes into account the quantum-physical nature of the carrier of information, channel, and the receiver measurement. The main accomplishment of HSW\ was to provide a mathematical specification of a decoding measurement that a receiver, bound only by the laws of quantum mechanics, could perform on the output codeword to recover the classical data transmitted by a sender at any rate below the Holevo limit. The HSW decoder prescription in general leads to a collective measurement on the codeword's joint quantum state, which may not be doable by detecting each individual symbol of the codeword separately.

For the single-mode lossy bosonic channel---which can be used to construct a wide class of practical free-space and fiber optical channels---it was shown that the single-letter Holevo bound is in fact the ultimate channel capacity~\cite{GGLMSY04}, given by%
\begin{equation}
g\left(  \eta N_{S}\right)  \equiv\left(  \eta N_{S}+1\right)  \log\left(
\eta N_{S}+1\right)  -\eta N_{S}\log\left(  \eta N_{S}\right)
\label{eq:capacity_lossybosonic}
\end{equation}
bits per channel use, where $N_{S}$ is the mean transmitted photon number per channel use, and $\eta \in (0, 1]$ is the input-output power transmissivity. Furthermore, conventional laser-light (coherent-state) modulation with symbols chosen i.i.d. from an isotropic Gaussian prior distribution, can achieve this capacity (i.e., it is not necessary to use exotic non-classical states, such as squeezed or entangled states). The lossy bosonic channel preserves a coherent state ($|\alpha\rangle \to |\sqrt{\eta}\alpha\rangle$), thus preserving its purity. The average output state is a zero-mean circularly-symmetric Gaussian mixture of coherent states, which is a thermal state with mean photon number $\eta N_{S}$, which saturates the entropy bound $g\left(\eta N_{S}\right)$. A converse proof shows that no other choice of modulation states and/or priors can exceed this capacity~\cite{GGLMSY04}. This result enabled comparing the ultimate channel capacity with the ideal Shannon limits of the classical channels induced by the quantum noise-characteristics of standard optical receivers, such as homodyne, heterodyne and direct detection receivers~\cite{GGLMSY04}. In spite of this accomplishment, it remains unclear how one could construct an implementation of the HSW decoding measurement for the bosonic channel using known optical components.

The theory of HSW also applies in the setting of \textquotedblleft quantum
reading\textquotedblright\ \cite{P11}, where one can obtain a quantum
advantage in the rate of read out of classical information stored in a digital memory.
Classical bits are encoded into the reflectivity and phase of memory cells. A transmitter irradiates the memory with light that in turn is modulated by a passive linear reflection from the memory cells (each cell is a single-mode lossy bosonic channel, but this time information is encoded in the memory cell's transmissivity and phase). A mono-static receiver gathers the reflected light for measurement and processing. The above is a bare-bone model for optical disks such as CDs or DVDs. Pirandola originally considered this task in the context of quantum channel discrimination and demonstrated a quantum advantage. He and his collaborators later considered a coded strategy (in the information-theoretic sense)~\cite{PLGMB11}. Later work \cite{GDNSY11,G12}\ improved upon Ref.~\cite{PLGMB11}, by demonstrating how to achieve the Holevo limit $g\left(N_{S}\right)$ bits/cell, where $N_{S}$ is the mean number of photons available at the transmitter to shine on each memory cell on an average. It turns out however, that the strategy for achieving $g\left(N_{S}\right)$ is different from that of the lossy bosonic channel, and surprisingly, a coherent-state probe fails to achieve the Holevo capacity~\cite{GDNSY11,G12}. The classical information is encoded into the phase of the cells (with each having perfect reflectivity). The symbols of the phase code are chosen i.i.d. and uniformly at random from the interval $[0,2\pi)$. The transmitter shines each cell with the single-mode quantum superposition state:%
\begin{equation}
|\phi_{\text{II}}\rangle\equiv\sum_{n=0}^{\infty}\sqrt{N_{S}^{n}/\left(
N_{S}+1\right)  ^{n+1}}\left\vert n\right\rangle ,\label{eq:Type-II-state}%
\end{equation}
and the receiver performs a collective measurement on the received codeword ($\left\vert n\right\rangle $ is a photon number state~\cite{GK04}). The average state of the received ensemble is a completely de-phased version of $|\phi_{\text{II}}\rangle$, yet again, a thermal state with mean photon number $N_{S}$, which saturates the entropy bound $g\left(N_{S}\right)$. Again, the authors of Ref.~\cite{GDNSY11} left open the question of a structured capacity-achieving receiver measurement.

In this paper, we address the open questions from Refs.~\cite{GGLMSY04,GDNSY11,G12}, by detailing a structured quantum measurement that can achieve both of the
above Holevo limits (for optical communication and quantum reading). The measurement is a sequential decoder, in the sense that it is a sequence of binary-outcome measurements that ask, \textquotedblleft Was the received quantum state
produced from the first codeword? the second codeword?\ the
third?\textquotedblright\ etc.,\ proceeding until the answer to one of the
questions is \textquotedblleft yes.\textquotedblright\ Our
work builds on recent insights of Giovannetti \textit{et al}.~\cite{GLM10} and
Sen~\cite{S11} in sequential decoding for quantum channels. Our primary contribution here is to show how to construct these measurements in an optical setting.

Our sequential decoding scheme for the lossy bosonic channel requires two
capabilities at the receiver. First, the receiver should be able to apply a
\textquotedblleft displacement operator,\textquotedblright\ which simply
requires highly reflective beamsplitters and a strong laser local
oscillator~\cite{P96}. Second, the receiver should be able to
perform a quantum non-demolition measurement to determine whether an $n$-mode
state is in the vacuum state or not. That is, the measurement operators are of
the form $\{\left\vert 0\right\rangle \left\langle 0\right\vert ^{\otimes
n},I^{\otimes n}-\left\vert 0\right\rangle \left\langle 0\right\vert ^{\otimes
n}\}$, where $\left\vert 0\right\rangle $ is the vacuum state and $I$ is the
identity operator. After performing such a measurement on an $n$-mode state
$\left\vert \psi\right\rangle $, the post-measurement state should be either
$\left\vert 0\right\rangle ^{\otimes n}$ or $(\left\vert
\psi\right\rangle -c\left\vert 0\right\rangle ^{\otimes n})/\sqrt{1-|c|^2}$, with
$c=\langle{0|^{\otimes n}|\psi}\rangle$. The key aspect of this
measurement is that its disturbance to an $n$-mode state
becomes asymptotically negligible as $n$ becomes large, as long as the
number of codewords is no larger than $\sim 2^{n g\left(\eta N_{S}\right)}$. Our sequential decoding
scheme for quantum reading requires the \textquotedblleft
vacuum-or-not\textquotedblright\ measurement described above, and the ability
to perform phase shifting and online squeezing \cite{GK04}.

We structure this paper as follows. Section~\ref{sec:def} reviews standard definitions and notation that are helpful for understanding the rest of the paper. 
Section~\ref{sec:sequential-decoding-proof} describes how a sequential
decoder operates when decoding classical information transmitted over a
pure-state classical-quantum channel, and for completeness, Appendix~\ref{sec:err-analysis} provides a proof that this scheme achieves the Holevo capacity.
Section~\ref{sec:sequential-dec-lossy-bosonic}\ provides a summary of the
operations needed for sequential decoding of the lossy bosonic channel.
Section~\ref{sec:sequential-decoder-q-reading}\ details an implementation of a
sequential decoder for quantum reading. We conclude in Section~\ref{sec:concl}%
\ with a summary and a list of open questions.

\section{Definitions and Notation}\label{sec:def}

We denote quantum systems as $A$, $B$, and $C$ and their corresponding Hilbert
spaces as $\mathcal{H}^{A}$, $\mathcal{H}^{B}$, and $\mathcal{H}^{C}$ with
respective dimensions $d_{A}$, $d_{B}$, and $d_{C}$. We denote pure states of
the system $A$ with a \emph{ket} $\left\vert \phi\right\rangle ^{A}$ and the
corresponding density operator as $\phi^{A}=\left\vert \phi\right\rangle
\!\left\langle \phi\right\vert ^{A}$. All kets that are quantum states have
unit norm, and all density operators are positive semi-definite with unit
trace. We model our lack of access to a quantum system with the partial trace
operation. That is, given a two-qubit state $\rho^{AB}$ shared between Alice
and Bob, we can describe Alice's state with the reduced density operator:
$\rho^{A}=$Tr$_{B}\left\{  \rho^{AB}\right\}  $, where Tr$_{B}$ denotes a
partial trace over Bob's system. Let $H(A)_{\rho}\equiv-$Tr$\left\{  \rho
^{A}\log\rho^{A}\right\}  $ be the von Neumann entropy of the state $\rho^{A}$.

\section{Sequential Decoding}

\label{sec:sequential-decoding-proof}In this section, we
describe the operation of a sequential decoder that can reliably recover classical information
encoded into a pure state ensemble. Appendix~\ref{sec:err-analysis} contains a full error analysis,
demonstrating that the scheme achieves capacity.

Suppose that a classical-quantum channel of the form $x\rightarrow\left\vert
\phi_{x}\right\rangle $ connects a sender Alice to a receiver Bob. For our
purposes here, it does not matter whether the classical input~$x$ is discrete
or continuous.

\begin{theorem}
\label{thm:seq-decoder}Let $x\rightarrow\left\vert \phi_{x}\right\rangle $ be
a classical-quantum channel and let $\rho\equiv\sum_{x}p_{X}\left(  x\right)
\left\vert \phi_{x}\right\rangle \left\langle \phi_{x}\right\vert $ for some
distribution $p_{X}\left(  x\right)  $. Then the rate $H\left(  \rho\right)$ bits per channel use is achievable for communication over this channel by having the receiver employ a
sequential decoding strategy.
\end{theorem}

\begin{proof}
We break the proof into several steps.

\textbf{Codebook Construction.} Before communication begins, Alice and Bob
agree upon a codebook. We allow them to select a codebook
randomly according to the distribution $p_{X}\left(  x\right)  $. So, for
every message $m\in\mathcal{M} \equiv \left\{1, \ldots, 2^{nR}\right\}$, generate a codeword $x^{n}\left(  m\right)
\equiv x_{1}\left(  m\right)  \cdots x_{n}\left(  m\right)  $ randomly and
independently according to%
\[
p_{X^{n}}\left(  x^{n}\right)  \equiv\prod\limits_{i=1}^{n}p_{X}\left(
x_{i}\right)  .
\]

\textbf{Sequential Decoding.} Transmitting the codeword $x^{n}\left(
m\right)  $ through $n$ uses of the channel $x\rightarrow\left\vert \phi
_{x}\right\rangle $ leads to the following quantum state at Bob's output:%
\[
\left\vert \phi_{x^{n}\left(  m\right)  }\right\rangle \equiv\left\vert
\phi_{x_{1}\left(  m\right)  }\right\rangle \otimes\cdots\otimes\left\vert
\phi_{x_{n}\left(  m\right)  }\right\rangle .
\]
Upon receiving the quantum codeword $\left\vert \phi_{x^{n}\left(  m\right)
}\right\rangle $, Bob performs a sequence of binary-outcome quantum
measurements to determine the classical codeword $x^{n}\left(  m\right)  $
that Alice transmitted. He first \textquotedblleft asks,\textquotedblright%
\ \textquotedblleft Is it the first codeword?\textquotedblright\ by performing
the measurement $\{\left\vert \phi_{x^{n}\left(  1\right)  }\right\rangle
\left\langle \phi_{x^{n}\left(  1\right)  }\right\vert ,I^{\otimes
n}-\left\vert \phi_{x^{n}\left(  1\right)  }\right\rangle \left\langle
\phi_{x^{n}\left(  1\right)  }\right\vert \}$. If he receives the outcome
\textquotedblleft yes,\textquotedblright\ then he performs no further
measurements and concludes that Alice transmitted the codeword $x^{n}\left(
1\right)  $. If he receives the outcome \textquotedblleft
no,\textquotedblright\ then he performs the measurement $\{\left\vert
\phi_{x^{n}\left(  2\right)  }\right\rangle \left\langle \phi_{x^{n}\left(
2\right)  }\right\vert ,I^{\otimes n}-\left\vert \phi_{x^{n}\left(  2\right)
}\right\rangle \left\langle \phi_{x^{n}\left(  2\right)  }\right\vert \}$ to
check if Alice sent the second codeword. Similarly, he stops if he receives
\textquotedblleft yes,\textquotedblright\ and otherwise, he proceeds along similar lines.

The above concludes the description of the operation of the sequential decoder.
We provide an error analysis demonstrating that this scheme works well in Appendix~\ref{sec:err-analysis}, i.e., the word error goes to zero as $n \to \infty$, as long as $R < H(\rho)$. Note that Sen \cite{S11} and Giovannetti \textit{et al}.~\cite{GLM10}\ already gave a proof that a sequential decoder works, but our proof in Appendix~\ref{sec:err-analysis} is a bit simpler because it is specialized to the case of pure-state ensembles (which is sufficient to consider for our settings of pure-loss optical communication and quantum reading).

\section{Sequential Decoding for Optical Communication}

\label{sec:sequential-dec-lossy-bosonic}We now provide a physical realization
of the sequential decoding strategy in the context of optical communications.
In this setting, we suppose that a lossy bosonic channel, specified by the
following Heisenberg relations, connects Alice to Bob:%
\begin{equation}
\hat{b}=\sqrt{\eta}\hat{a}+\sqrt{1-\eta}\hat{e}%
,\label{eq:lossy-bosonic-channel}%
\end{equation}
where $\hat{a}$, $\hat{b}$, and $\hat{e}$ are the respective field operators
for Alice's input mode, Bob's output mode, and an environmental input mode
(assumed to be in its vacuum state). The transmissivity $\eta\in\left[  0,1\right]$ is the fraction of Alice's input photons that make it to Bob on average. We assume that Alice is
constrained to using mean photon number $N_{S}$ per channel use.

The strategy for achieving the classical capacity of this channel is for Alice
to induce a classical-quantum channel, by selecting $\alpha \in {\mathbb C}$ and preparing a coherent state $\left\vert \alpha\right\rangle $
\cite{GK04}\ at the input of the channel in (\ref{eq:lossy-bosonic-channel}).
The resulting induced classical-quantum channel to Bob is of the following
form:%
\[
\alpha\rightarrow|\sqrt{\eta}\alpha\rangle.
\]
By choosing the distribution $p_{X}\left(  x\right)  $
in Theorem~\ref{thm:seq-decoder}\ to be an isotropic, complex Gaussian with
variance $N_{S}$:%
\[
p_{N_{S}}\left(  \alpha\right)  \equiv \left({1}/{\pi N_{S}}\right)\exp\left\{
{-\left\vert \alpha\right\vert ^{2}}/{N_{S}}\right\}  ,
\]
we have that $g\left(  \eta N_{S}\right)  $ is
an achievable rate for classical communication. The quantity $g\left(  \eta
N_{S}\right)  $ is the entropy of the average state of the ensemble
$\{p_{N_{S}}\left(  \alpha\right)  ,\ |\sqrt{\eta}\alpha\rangle\}$:%
\[
\int d^{2}\alpha\ p_{N_{S}}\left(  \alpha\right)  |\sqrt{\eta}\alpha
\rangle\langle\sqrt{\eta}\alpha|,
\]
which is a thermal state with mean photon number $\eta N_{S}$ \cite{GK04}.

Each quantum codeword selected from the ensemble $\{p_{N_{S}}\left(
\alpha\right)  ,\ |\alpha\rangle\}$ has the following form:%
\[
\left\vert \alpha^{n}\left(  m\right)  \right\rangle \equiv\left\vert
\alpha_{1}\left(  m\right)  \right\rangle \otimes\cdots\otimes\left\vert
\alpha_{n}\left(  m\right)  \right\rangle .
\]
We assume $\eta=1$ above and for the rest of this section without loss of generality. Thus, the sequential decoder consists of measurements of the following form for all
$m\in\mathcal{M}$:%
\begin{equation}
\left\{  \left\vert \alpha^{n}\left(  m\right)  \right\rangle \left\langle
\alpha^{n}\left(  m\right)  \right\vert ,\ I^{\otimes n}-\left\vert \alpha
^{n}\left(  m\right)  \right\rangle \left\langle \alpha^{n}\left(  m\right)
\right\vert \right\}  .\label{eq:seq-decoder-bosonic}%
\end{equation}
Observing that%
\[
\left\vert \alpha^{n}\left(  m\right)  \right\rangle =D\left(  \alpha
_{1}\left(  m\right)  \right)  \otimes\cdots\otimes D\left(  \alpha_{n}\left(
m\right)  \right)  \left\vert 0\right\rangle ^{\otimes n},
\]
where $D\left(  \alpha\right)  \equiv\exp\left\{  \alpha\hat{a}^{\dag}%
-\alpha^{\ast}\hat{a}\right\}  $ is the well-known unitary \textquotedblleft
displacement\textquotedblright\ operator from quantum optics \cite{GK04}\ and
$\left\vert 0\right\rangle ^{\otimes n}$ is the $n$-fold tensor product vacuum
state, it is clear that the decoder can implement the measurement in
(\ref{eq:seq-decoder-bosonic}) in three steps:

\begin{enumerate}
\item Displace the $n$-mode codeword state by%
\begin{equation*}
D\left(  -\alpha_{1}\left(  m\right)  \right)  \otimes\cdots\otimes D\left(
-\alpha_{n}\left(  m\right)  \right)  ,
\end{equation*}
by employing highly asymmetric beam-splitters with a strong local oscillator \cite{P96}.

\item Perform a \textquotedblleft vacuum-or-not\textquotedblright\ measurement
of the form%
\[
\left\{  \left\vert 0\right\rangle \left\langle 0\right\vert ^{\otimes
n},\ \ I^{\otimes n}-\left\vert 0\right\rangle \left\langle 0\right\vert
^{\otimes n}\right\}  .
\]
If the vacuum outcome occurs, decode as the $m^{\text{th}}$ codeword.
Otherwise, proceed.

\item Displace by $D\left(  \alpha_{1}\left(  m\right)  \right)  \otimes
\cdots\otimes D\left(  \alpha_{n}\left(  m\right)  \right)  $ with the same
method as in Step 1.
\end{enumerate}

The receiver just iterates this strategy for every codeword in the codebook,
and Theorem~\ref{thm:seq-decoder}\ states this strategy is capacity-achieving.

\begin{remark}
The above strategy is reminiscent of the class of conditional pulse nulling
receivers \cite{GHT10}, which are useful in discriminating $M$-ary
pulse-position-modulation coded states with $\left\vert \alpha\right\rangle $
in the $i^{\text{th}}$ slot and vacuum states $\left\vert 0\right\rangle $ in
the other $M-1$ slots. In this strategy, the receiver hypothesizes at first
that the transmitted codeword is the first codeword $\left\vert \alpha
\right\rangle \left\vert 0\right\rangle ^{\otimes M-1}$, nulls the first mode
by applying $D^{\dag}\left(  \alpha\right)  $, and direct-detects the first
mode. If the sender in fact transmitted the first codeword, then the resulting
state is ideally $\left\vert 0\right\rangle ^{\otimes m}$, and direct
detection of the first mode should ideally produce no \textquotedblleft
clicks.\textquotedblright\ If there is no click, then the receiver direct
detects the other modes to confirm the original hypothesis. If there are no
further clicks, then the receiver declares that the sender transmitted the
first codeword. If there is a further click, then the receiver
guesses the codeword corresponding to the position of the click. If on the first mode there is a click, then the receiver hypothesizes that the transmitted codeword is the second one and repeats the above algorithm on the next $M-1$ modes.

The difference between the sequential decoding strategy and conditional pulse
nulling is that the codewords are different, and the vacuum-or-not measurement
in the sequential decoding strategy is much more difficult to perform in
practice than direct detection, which annihilates the detected quantum state. Ideally, the vacuum-or-not should be a non-demolition measurement such that the post-measurement state is $\left\vert
0\right\rangle ^{\otimes n}$ or $(\left\vert \psi\right\rangle -c\left\vert
0\right\rangle ^{\otimes n})/\sqrt{1-|c|^2}$, with $c=\langle{0|^{\otimes n}|\psi}\rangle$, if the pre-measurement state is $\left\vert \psi\right\rangle $, with probabilities $p_0 = \left|c\right|^2$ and $p_1=1-p_0$, respectively, of the two possible outcomes.
\end{remark}

\begin{remark}
The crucial (and most difficult) step in sequential decoding for the lossy
bosonic channel is the vacuum-or-not measurement. Oi \textit{et al}.~have
provided a method for performing this measurement, by interacting the light
field with a three-level atom in a STIRAP\ process \cite{Oi2012}. This approach
would likely be quite lossy in practice, so it would be ideal to determine an
all-optical vacuum-or-not measurement.
\end{remark}

\begin{remark}
If the mean input photon number $N_{S} \ll 1$, then one does not require a
full Gaussian distributed codebook in order to achieve capacity. A simpler
method, called binary phase-shift keying, suffices to approach capacity very closely. In
this approach, the ensemble for generating a codebook randomly is just
$\left\{  1/2,\ \left\vert \pm\alpha\right\rangle \right\}  $. This also
simplifies the sequential decoder because the only displacements required for
implementation are $D\left(  \pm\alpha\right)  $. An additional advantage is that a random
linear encoder should achieve the capacity, by an argument similar to that on pages
3-14 and 3-15 of Ref.~\cite{el2010lecture}. BPSK polar codes are capacity-achieving for low-photon
number as well \cite{WG11}.
\end{remark}

\begin{remark}
Tan proved a variation of Theorem~\ref{thm:seq-decoder}\ for the lossy bosonic
channel in her thesis \cite{T10}, but the analysis in Appendix~\ref{sec:err-analysis} demonstrates that it is actually not necessary
to perform a measurement onto the average typical subspace. We avoided having
to do so by demonstrating that it is sufficient to code for a
typical-projected version of the channel and applying Sen's non-commutative
union bound from Ref.~\cite{S11}.
\end{remark}

\begin{remark}
The above sequential decoding approach also works well in the context of
private classical communication over a lossy bosonic channel \cite{D05,GSE08}.
The private classical capacity of the channel in
(\ref{eq:lossy-bosonic-channel}) is $g\left(  \eta N_{S}\right)  -g\left(
\left(  1-\eta\right)  N_{S}\right)  $ (compare to its public classical
capacity of $g\left(  \eta N_{S}\right)  $), and the strategy for encoding is
again to choose coherent states randomly according to an isotropic Gaussian
prior. The sequential decoder can just test for all codewords in a codebook of
size $2^{ng\left(  \eta N_{S}\right)  }$ and recover the transmitted private
message correctly. The privacy in the scheme comes about by choosing
$2^{ng\left(  \left(  1-\eta\right)  N_{S}\right)  }$ codewords corresponding
to each message and selecting one of these uniformly at random in order to
randomize Eve's knowledge of the transmitted message \cite{D05}.
\end{remark}

\section{Sequential Decoding for Quantum Reading}

\label{sec:sequential-decoder-q-reading}The sequential decoding strategy also
finds application in \textquotedblleft quantum reading\textquotedblright%
\ \cite{P11}.\ In this setting, we suppose that information is encoded into
passive memory cells of an optically-readable memory, which a transceiver can read out by irradiating them with laser (or quantum) light and detecting the reflected light. More specifically, we can model the $i^{\text{th}}$ optical memory cell as a beamsplitter of the following form:%
\[
\hat{b}_{i}=\exp\{\text{i}\theta_{i}\}\sqrt{\eta_{i}}\hat{a}_{i}%
+\sqrt{1-\eta_{i}}\hat{e}_{i},
\]
where the parameters $\eta_{i}$ and $\theta_{i}$ are the respective
reflectivity and phase of the $i^{\text{th}}$ cell, and $\hat{a}_{i}$,
$\hat{b}_{i}$, and $\hat{e}_{i}$ are the respective field operators for the
transmitter's $i^{\text{th}}$ input mode, the $i^{\text{th}}$ reflected mode,
and an environmental mode (assumed to be in its vacuum state). We assume perfect channels from the transmitter to the optical memory cells and from the cells back to the receiver (which is co-located with the transmitter).

The objective is for the transmitter to interrogate each optical memory cell
with some quantum state of light with mean photon number $N_{S}$. The receiver
then collects all of the reflected light and performs some measurement to
recover the classical information encoded in the memory cells. If we use a coherent-state transmitter to interrogate each cell, we call it the Type I setting~ \cite{GDNSY11}. If we do not allow the transmitter to retain any state entangled with the transmitted light, but allow it to send any quantum state (entangled spatially across modes or an unentangled non-classical product state), then this is termed the Type II\ setting \cite{GDNSY11}. Finally, if we do allow for entanglement assistance, in the sense that the transmitter can prepare two modes in an entangled state for each of the $n$~memory cells, send one to a memory cell while retaining the
other, then this is termed a Type III\ setting~\cite{GDNSY11}. In each of the three settings, the receiver is always allowed to perform a general (collective) quantum measurement on the reflected $n$ modes (and the retained $n$ modes, in case of Type III). It is straightforward to prove that $g\left(  N_{S}\right)  $
is the Holevo (upper) bound on the capacity of quantum reading in the Type I\ and Type II\ settings, while it is unknown whether $g\left(  N_{S}\right)$ could be exceeded 
in the Type III setting~\cite{GDNSY11}.

Recently, Guha \textit{et al}.~proved that the following strategy achieves the
$g\left(  N_{S}\right)$ bound for quantum reading using a Type II transmitter \cite{GDNSY11,G12}. The transmitter interrogates each memory cell with a quantum state of light of the form in (\ref{eq:Type-II-state}). It is straightforward to compute that the
mean number of photons in this state is $N_{S}$:\ $\langle\phi_{\text{II}%
}|\ \hat{n}\ |\phi_{\text{II}}\rangle=N_{S}$, where $\hat{n}={\hat a}^\dagger{\hat a}$ is the photon number operator \cite{GK04}. Each memory cell has classical information
encoded into only the phase variable $\theta_{i}$ (with $\eta_{i}=1$), so that
a randomly chosen code in the sense of Theorem~\ref{thm:seq-decoder}\ is
selected from the following ensemble:%
\begin{equation}
\left\{  1/2\pi,\ |\phi_{\text{II},\theta}\rangle\right\}
,\label{eq:phase-ensemble}%
\end{equation}
where%
\begin{equation}
|\phi_{\text{II},\theta}\rangle\equiv\sum_{n=0}^{\infty}\sqrt{N_{S}^{n}/\left(
N_{S}+1\right)  ^{n+1}}\exp\{\text{i}n\theta\}\left\vert n\right\rangle,\label{eq:phase-shifted-Type-II}%
\end{equation}
and each $\theta$ is selected uniformly at random from the interval $[0,2\pi
)$. The average state of this code ensemble is%
\[
\frac{1}{2\pi}\int\limits_{0}^{2\pi}d\theta\ |\phi_{\text{II},\theta}%
\rangle\langle\phi_{\text{II},\theta}|\ =\sum_{n=0}^{\infty}N_{S}^{n}/\left(
N_{S}+1\right)  ^{n+1}\left\vert n\right\rangle \left\langle n\right\vert ,
\]
which is a thermal state with mean photon number $N_{S}$. (The effect of
phase-randomizing the state $|\phi_{\text{II}}\rangle$ is simply to de-phase it
to a thermal state.) Thus, a random code constructed from the ensemble in
(\ref{eq:phase-ensemble}) along with a sequential decoder saturates the
entropy bound $g\left(N_{S}\right)  $ because the average state is a thermal state.

It is not clear to us at the moment how to implement a sequential decoder for
the above Type II\ strategy. Though, if we allow for a Type III\ transmitter,
the strategy is straightforward to specify. First, the transmitter
interrogates each optical memory cell with one mode of a two-mode squeezed
vacuum state \cite{GK04}\ of the following form:%
\[
|\phi_{\text{III}}\rangle\equiv\sum_{n=0}^{\infty}\sqrt{N_{S}^{n}/\left(
N_{S}+1\right)  ^{n+1}}\left\vert n\right\rangle \left\vert n\right\rangle ,
\]
while retaining the other mode. The encoding in the optical memory cells is
the same as above, such that the memory cells have classical information
encoded only into a uniformly random phase. The code ensemble is then
$\{1/2\pi,\ |\phi_{\text{III},\theta}\rangle\}$, with $|\phi_{\text{III}%
,\theta}\rangle$ defined similarly as in (\ref{eq:phase-shifted-Type-II}). The authors of Ref.~\cite{GDNSY11} showed that this ensemble also saturates the $g(N_S)$ bound.
Consider the $m^{\text{th}}$ quantum codeword to have the form:%
\[
|\phi_{\text{III},\theta^{n}\left(  m\right)  }\rangle\equiv|\phi
_{\text{III},\theta_{1}\left(  m\right)  }\rangle\otimes\cdots\otimes
|\phi_{\text{III},\theta_{n}\left(  m\right)  }\rangle.
\]
Consider further that each of the states in the above tensor product can be
written as%
\[
|\phi_{\text{III},\theta_{i}\left(  m\right)  }\rangle=\left(  P\left(
\theta_{i}\left(  m\right)  \right)  \otimes I\right)  S\left(  r\right)
\left\vert 0\right\rangle ^{\otimes2},
\]
where $P\left(  \theta_{i}\left(  m\right)  \right)  =\exp\{\text{i}\hat{n}\theta_{i}\left(  m\right)\}
$ is a phase shifter, $S\left(  r\right)  $ is a two-mode squeezing
operator \cite{GK04}\ with the squeezing strength $r$, s.t. $N_{S} = \sinh^2r$,
and $\left\vert 0\right\rangle ^{\otimes2}$ is a two-mode vacuum state. This
then leads us to specify the $m^{\text{th}}$ step of the sequential decoder,
which proceeds as follows:

\begin{enumerate}
\item Apply the operator $\left(  P^{\dag}\left(  \theta_{i}\left(  m\right)
\right)  \otimes I\right)  $ by phase-shifting the first mode of the
$i^{\text{th}}$ pair by $-\theta_{i}\left(  m\right)  $.

\item Apply the unsqueezing operator $S^{\dag}\left(  r\right)  $. The
receiver can accomplish this with a phase-sensitive amplifier.

\item Perform a \textquotedblleft vacuum-or-not\textquotedblright\ measurement
of the same form as in Step~2 in the previous section.
If the vacuum outcome occurs, decode as the $m^{\text{th}}$ codeword.
Otherwise, proceed.

\item Apply the squeezing operator $S\left(  r\right)  $.

\item Apply the operator $\left(  P\left(  \theta_{i}\left(  m\right)
\right)  \otimes I\right)  $ by phase-shifting the first mode of the
$i^{\text{th}}$ pair by $\theta_{i}\left(  m\right)  $.
\end{enumerate}

The receiver again iterates this strategy for all codewords in the codebook,
and Theorem~\ref{thm:seq-decoder}\ states that this strategy is Holevo-capacity-achieving, i.e., it achieves $g(N_S)$ bits/cell.

\begin{remark}
At $N_S \ll 1$, a binary phase-shift keying code approximately achieves
the Holevo limit of $g\left(N_{S}\right)$, i.e., $C_{\rm BPSK}(N_S) = H_2\left((1 \pm e^{-2N_S})/2\right)$. The code ensemble for this case
is just $\left\{  \left(  1/2,|\phi_{\text{III}}\rangle\right)  ,\left(
1/2,|\phi_{\text{III},\pi}\rangle\right)  \right\}  $, and the sequential
decoder only needs to have phase shifts of 0 or $\pi$. Interestingly enough, with binary phase modulation, even a coherent-state (Type I) transmitter can achieve $C_{\rm BPSK}(N_S)$.
\end{remark}

\section{Conclusion}

\label{sec:concl}We have demonstrated that a sequential decoding strategy
achieves the Holevo capacity for optical communication and quantum reading, by building on information-theoretic works on sequential decoding in Refs.~\cite{GLM10,S11}. Both schemes employ a \textquotedblleft vacuum-or-not\textquotedblright\ measurement which distinguishes coherently and in a non-demolition way between the vacuum or \textquotedblleft not
vacuum,\textquotedblright\ so that the disturbance on the encoded state is
asymptotically negligible for long codewords (as long as the code rate is less than the Holevo limit). For optical communication, the only other operation needed is implementing a displacement operator, while the sequential quantum reading
receiver requires phase shifting and online squeezing.

The most important open problems going forward concern making the scheme more practical. In this vein, it might be helpful to realize an all-optical implementation of the \textquotedblleft vacuum-or-not\textquotedblright\ measurement---which could help both on the scalability front, and relative ease of implementation as compared to a system that uses atom-light interaction~\cite{Oi2012}. Also, the sequential decoding scheme given here is impractical from a computational perspective because it requires an exponential number of measurements (there are an exponential number of codewords). It would be better to have a sequential decoder that decodes one bit at a time and would thus require only a linear number of measurements. The polar decoder for classical-quantum channels is one such sequential decoder~\cite{WG11}, but it remains unclear to us how to implement it with optical devices.

We thank J. P.~Dowling, V. Giovannetti, P. Hayden, L. Maccone, and J. H.~Shapiro for useful discussions.
\bibliographystyle{IEEEtran}
\bibliography{Ref}

\newpage
\newpage
\appendices

\section{Important Lemmas}

In order to describe the \textquotedblleft distance\textquotedblright\ between
two quantum states, we use the notion of \emph{trace distance}. The trace
distance between states $\sigma$ and $\rho$ is $\Vert\sigma-\rho\Vert
_{1}=\mathrm{Tr}\left\vert \sigma-\rho\right\vert $, where $\left\vert
X\right\vert =\sqrt{X^{\dagger}X}$. Two states that are similar have trace
distance close to zero, whereas states that are perfectly distinguishable have
trace distance equal to two.

Two states can substitute for one  another up to a penalty proportional to the
trace distance between them: 

\begin{lemma}
\label{lem:tr-trick}  Let  $0\leq\rho, \sigma, \Lambda\leq I$. Then
\begin{equation}
\mathrm{Tr}\left[  \Lambda\rho\right]  \leq\mathrm{Tr}\left[  \Lambda
\sigma\right]  + \left\Vert \rho-\sigma\right\Vert _{1}.\label{eqn:tr-trick}%
\end{equation}

\end{lemma}

\begin{IEEEproof}
	This follows from a variational characterization of
	trace distance as the distinguishability of
	the states under an optimal measurement $M$ \cite{W11}:
	$\left\Vert \rho-\sigma\right\Vert _{1} = 2
	\max_{0 \leq M \leq I} \mathrm{Tr}\left[ M(\rho-\sigma) \right]$.
	\end{IEEEproof}

Consider a density operator $\rho$ with the following spectral decomposition:%
\[
\rho=\sum_{x}p_{X}\left(  x\right)  \left\vert x\right\rangle \left\langle
x\right\vert .
\]
The weakly typical subspace is defined as the span of all vectors such that
the sample entropy $\overline{H}\left(  x^{n}\right)  $ of their classical
label is close to the true entropy $H\left(  X\right)  $ of the distribution
$p_{X}\left(  x\right)  $ \cite{W11}:%
\[
T_{\delta}^{X^{n}}\equiv\text{span}\left\{  \left\vert x^{n}\right\rangle
:\left\vert \overline{H}\left(  x^{n}\right)  -H\left(  X\right)  \right\vert
\leq\delta\right\}  ,
\]
where%
\begin{align*}
\overline{H}\left(  x^{n}\right)   &  \equiv-\frac{1}{n}\log\left(  p_{X^{n}%
}\left(  x^{n}\right)  \right)  ,\\
H\left(  X\right)   &  \equiv-\sum_{x}p_{X}\left(  x\right)  \log p_{X}\left(
x\right)  .
\end{align*}
The projector $\Pi_{\rho,\delta}^{n}$\ onto the typical subspace of $\rho$ is
defined as%
\[
\Pi_{\rho,\delta}^{n}\equiv\sum_{x^{n}\in T_{\delta}^{X^{n}}}\left\vert
x^{n}\right\rangle \left\langle x^{n}\right\vert ,
\]
where we have \textquotedblleft overloaded\textquotedblright\ the symbol
$T_{\delta}^{X^{n}}$ to refer also to the set of $\delta$-typical sequences:%
\[
T_{\delta}^{X^{n}}\equiv\left\{  x^{n}:\left\vert \overline{H}\left(
x^{n}\right)  -H\left(  X\right)  \right\vert \leq\delta\right\}  .
\]
The three important properties of the typical projector are as follows:%
\begin{align*}
\text{Tr}\left\{  \Pi_{\rho,\delta}^{n}\rho^{\otimes n}\right\}   &
\geq1-\epsilon,\\
\text{Tr}\left\{  \Pi_{\rho,\delta}^{n}\right\}   &  \leq2^{n\left[  H\left(
X\right)  +\delta\right]  },\\
2^{-n\left[  H\left(  X\right)  +\delta\right]  }\Pi_{\rho,\delta}^{n}  &
\leq\Pi_{\rho,\delta}^{n}\rho^{\otimes n}\Pi_{\rho,\delta}^{n}\leq2^{-n\left[
H\left(  X\right)  -\delta\right]  }\Pi_{\rho,\delta}^{n},
\end{align*}
where the first property holds for arbitrary $\epsilon,\delta>0$ and
sufficiently large $n$.

\begin{lemma}
[Gentle Operator Lemma for Ensembles]\label{lem:gentle-operator} Given an
ensemble $\left\{  p_{X}\left(  x\right)  ,\rho_{x}\right\}  $ with expected
density operator $\rho\equiv\sum_{x}p_{X}\left(  x\right)  \rho_{x}$, suppose
that an operator $\Lambda$ such that $I\geq\Lambda\geq0$ succeeds with high
probability on the state $\rho$:%
\[
\text{Tr}\left\{  \Lambda\rho\right\}  \geq1-\epsilon.
\]
Then the subnormalized state $\sqrt{\Lambda}\rho_{x}\sqrt{\Lambda}$ is close
in expected trace distance to the original state $\rho_{x}$:%
\[
\mathbb{E}_{X}\left\{  \left\Vert \sqrt{\Lambda}\rho_{X}\sqrt{\Lambda}%
-\rho_{X}\right\Vert _{1}\right\}  \leq2\sqrt{\epsilon}.
\]

\end{lemma}

A proof of the above lemma is available in Ref.~\cite{W11}.

\section{Error Analysis for Sequential Decoding}

\label{sec:err-analysis}

In general, if Alice transmits the $m^{\text{th}}$
codeword, then the probability for Bob to decode correctly with this
sequential decoding strategy is as follows:%
\[
\text{Tr}\left\{  \phi_{x^{n}\left(  m\right)  }\hat{\Pi}_{m-1}\cdots\hat{\Pi
}_{1}\phi_{x^{n}\left(  m\right)  }\hat{\Pi}_{1}\cdots\hat{\Pi}_{m-1}%
\phi_{x^{n}\left(  m\right)  }\right\}  ,
\]
where we make the abbreviations%
\begin{align*}
\phi_{x^{n}\left(  m\right)  }  &  \equiv\left\vert \phi_{x^{n}\left(
m\right)  }\right\rangle \left\langle \phi_{x^{n}\left(  m\right)
}\right\vert ,\\
\hat{\Pi}_{i}  &  \equiv I^{\otimes n}-\left\vert \phi_{x^{n}\left(  i\right)
}\right\rangle \left\langle \phi_{x^{n}\left(  i\right)  }\right\vert .
\end{align*}
So the probability that Bob makes an error when decoding the $m^{\text{th}}$
codeword is just%
\[
1-\text{Tr}\left\{  \phi_{x^{n}\left(  m\right)  }\hat{\Pi}_{m-1}\cdots
\hat{\Pi}_{1}\phi_{x^{n}\left(  m\right)  }\hat{\Pi}_{1}\cdots\hat{\Pi}%
_{m-1}\phi_{x^{n}\left(  m\right)  }\right\}  .
\]
To further simplify the error analysis, we consider the expectation of the
above error probability, under the assumption that Alice selects a message
uniformly at random according to a random variable $M$ and that the codeword
$x^{n}$ is selected at random according to the distribution $p_{X^{n}}\left(
x^{n}\right)  $ (as described above):%
\begin{equation}
1-\underset{X^{n},M}{\mathbb{E}}\text{Tr}\left\{  \phi_{X^{n}\left(  M\right)
}\hat{\Pi}_{M-1}\cdots\hat{\Pi}_{1}\phi_{X^{n}\left(  M\right)  }\hat{\Pi}%
_{1}\cdots\hat{\Pi}_{M-1}\right\}  . \label{eq:1st-err-prob}%
\end{equation}
For the rest of the proof, it is implicit that the expectation $\mathbb{E}$ is
with respect to random variables $X^{n}$ and $M$.

Our first observation is that, for the purposes of our error analysis, we can
\textquotedblleft smooth\textquotedblright\ the channel $x^{n}\rightarrow
\phi_{x^{n}}$, by imagining instead that we are coding for a projected version
of the channel $\Pi\ \phi_{x^{n}}\ \Pi$, where $\Pi$ is the typical projector
for the average state $\rho\equiv\sum_{x}p_{X}\left(  x\right)  \phi_{x}$.
Doing so simplifies the error analysis by cutting off large eigenvalues that
reside outside of the high-probability typical subspace. Furthermore, we
expect that doing so should not affect the error analysis very much because
most of the probability tends to concentrate in this subspace anyway. That we
can do so follows from the fact that%
\begin{align*}
1 &  =\mathbb{E}\text{Tr}\left\{  \phi_{X^{n}\left(  M\right)  }\right\}  \\
&  =\mathbb{E}\text{Tr}\left\{  \Pi\phi_{X^{n}\left(  M\right)  }\right\}
+\mathbb{E}\text{Tr}\left\{  \hat{\Pi}\phi_{X^{n}\left(  M\right)  }\right\}
\\
&  =\mathbb{E}\text{Tr}\left\{  \Pi\phi_{X^{n}\left(  M\right)  }\Pi\right\}
+\text{Tr}\left\{  \hat{\Pi}\mathbb{E}\phi_{X^{n}\left(  M\right)  }\right\}
\\
&  =\mathbb{E}\text{Tr}\left\{  \Pi\phi_{X^{n}\left(  M\right)  }\Pi\right\}
+\text{Tr}\left\{  \hat{\Pi}\rho^{\otimes n}\right\}  ,
\end{align*}
where $\hat{\Pi}\equiv I-\Pi$. Furthermore, we know that%
\begin{align*}
&  \mathbb{E}\text{Tr}\left\{  \phi_{X^{n}\left(  M\right)  }\hat{\Pi}%
_{M-1}\cdots\hat{\Pi}_{1}\phi_{X^{n}\left(  M\right)  }\hat{\Pi}_{1}\cdots
\hat{\Pi}_{M-1}\right\}  \\
&  =\mathbb{E}\text{Tr}\left\{  \hat{\Pi}_{1}\cdots\hat{\Pi}_{M-1}\phi
_{X^{n}\left(  M\right)  }\hat{\Pi}_{M-1}\cdots\hat{\Pi}_{1}\phi_{X^{n}\left(
M\right)  }\right\}  \\
&  \geq\mathbb{E}\text{Tr}\left\{  \hat{\Pi}_{1}\cdots\hat{\Pi}_{M-1}%
\phi_{X^{n}\left(  M\right)  }\hat{\Pi}_{M-1}\cdots\hat{\Pi}_{1}\Pi\phi
_{X^{n}\left(  M\right)  }\Pi\right\}  \\
&  \ \ \ \ \ \ \ \ \ \ \ -\mathbb{E}\left\Vert \phi_{X^{n}\left(  M\right)
}-\Pi\phi_{X^{n}\left(  M\right)  }\Pi\right\Vert _{1},
\end{align*}
where the inequality follows from Lemma~\ref{lem:tr-trick}. Using the above
observations and the facts that%
\begin{align}
\mathbb{E}\left\Vert \phi_{X^{n}\left(  M\right)  }-\Pi\phi_{X^{n}\left(
M\right)  }\Pi\right\Vert _{1} &  \leq2\sqrt{\epsilon}%
,\label{eq:trace-bound-tpy}\\
\text{Tr}\left\{  \hat{\Pi}\rho^{\otimes n}\right\}   &  \leq\epsilon,
\end{align}
for all $\epsilon>0$ whenever $n$ is sufficiently large (these are from the
properties of typicality and Lemma~\ref{lem:gentle-operator}), we obtain the
following upper bound on (\ref{eq:1st-err-prob}):%
\begin{multline}
\mathbb{E}\text{Tr}\left\{  \Pi\phi_{X^{n}\left(  M\right)  }\Pi\right\}
-\label{eq:error-prob-smoothed}\\
\mathbb{E}\text{Tr}\left\{  \phi_{X^{n}\left(  M\right)  }\hat{\Pi}%
_{M-1}\cdots\hat{\Pi}_{1}\Pi\phi_{X^{n}\left(  M\right)  }\Pi\hat{\Pi}%
_{1}\cdots\hat{\Pi}_{M-1}\phi_{X^{n}\left(  M\right)  }\right\}  \\
+\epsilon+2\sqrt{\epsilon}.
\end{multline}
(In the next steps, we omit the terms $\epsilon+2\sqrt{\epsilon}$ as they are
negligible.) The most important step of this error analysis is to apply Sen's
non-commutative union bound (Lemma~3 of Ref.~\cite{S11}), which holds for any
subnormalized state $\sigma$ ($\sigma\geq0$ and Tr$\left\{  \sigma\right\}
\leq1$) and sequence of projectors $\Pi_{1}$, \ldots, $\Pi_{N}$:%
\[
\text{Tr}\left\{  \sigma\right\}  -\text{Tr}\left\{  \Pi_{N}\cdots\Pi
_{1}\sigma\Pi_{1}\cdots\Pi_{N}\right\}  \leq2\sqrt{\sum_{i=1}^{N}%
\text{Tr}\left\{  \left(  I-\Pi_{i}\right)  \sigma\right\}  }%
\]
For our case, we take $\Pi\phi_{X^{n}\left(  M\right)  }\Pi$ as $\sigma$ and
$\phi_{X^{n}\left(  M\right)  }$, $\hat{\Pi}_{M-1}$, \ldots, $\hat{\Pi}_{1}$
as the sequence of projectors. Applying Sen's bound and concavity of the
square root function leads to the following upper bound on
(\ref{eq:error-prob-smoothed}):%
\[
2\sqrt{\mathbb{E}\text{Tr}\left\{  \hat{\Pi}_{M}\Pi\phi_{X^{n}\left(
M\right)  }\Pi\right\}  +\mathbb{E}\sum_{i=1}^{M-1}\text{Tr}\left\{
\phi_{X^{n}\left(  i\right)  }\Pi\phi_{X^{n}\left(  M\right)  }\Pi\right\}  }%
\]
where $\hat{\Pi}_{M}=I^{\otimes n}-\phi_{X^{n}\left(  M\right)  }$ and
$\phi_{X^{n}\left(  i\right)  }=I^{\otimes n}-\hat{\Pi}_{i}$. We now bound
each of the above two terms individually. For the first term, consider that%
\begin{align*}
&  \mathbb{E}\text{Tr}\left\{  \hat{\Pi}_{M}\Pi\phi_{X^{n}\left(  M\right)
}\Pi\right\}  \\
&  \leq\mathbb{E}\text{Tr}\left\{  \hat{\Pi}_{M}\phi_{X^{n}\left(  M\right)
}\right\}  +\mathbb{E}\left\Vert \phi_{X^{n}\left(  M\right)  }-\Pi\phi
_{X^{n}\left(  M\right)  }\Pi\right\Vert _{1}\\
&  \leq2\sqrt{\epsilon}.
\end{align*}
where the last inequality follows from applying (\ref{eq:trace-bound-tpy}) and
because%
\begin{align*}
\text{Tr}\left\{  \hat{\Pi}_{M}\phi_{X^{n}\left(  M\right)  }\right\}   &
=\text{Tr}\left\{  \left(  I^{\otimes n}-\phi_{X^{n}\left(  M\right)
}\right)  \phi_{X^{n}\left(  M\right)  }\right\}  \\
&  =0.
\end{align*}
For the second term, consider that%
\begin{align*}
&  \mathbb{E}\sum_{i=1}^{M-1}\text{Tr}\left\{  \phi_{X^{n}\left(  i\right)
}\Pi\phi_{X^{n}\left(  M\right)  }\Pi\right\}  \\
&  \leq\mathbb{E}_{M}\sum_{i\neq M}\mathbb{E}_{X^{n}}\text{Tr}\left\{
\phi_{X^{n}\left(  i\right)  }\Pi\phi_{X^{n}\left(  M\right)  }\Pi\right\}  \\
&  =\mathbb{E}_{M}\sum_{i\neq M}\text{Tr}\left\{  \mathbb{E}_{X^{n}}\left\{
\phi_{X^{n}\left(  i\right)  }\right\}  \Pi\mathbb{E}_{X^{n}}\left\{
\phi_{X^{n}\left(  M\right)  }\right\}  \Pi\right\}  \\
&  =\sum_{i\neq M}\text{Tr}\left\{  \rho^{\otimes n}\Pi\rho^{\otimes n}%
\Pi\right\}  \\
&  \leq2^{-n\left[  H\left(  \rho\right)  -\delta\right]  }\sum_{i\neq
M}\text{Tr}\left\{  \rho^{\otimes n}\Pi\right\}  \\
&  \leq2^{-n\left[  H\left(  \rho\right)  -\delta\right]  }\ \left\vert
\mathcal{M}\right\vert
\end{align*}
The first inequality follows by just adding in all of the future terms $i>M$
to the sum. The first equality follows because the random variables
$X^{n}\left(  i\right)  $ and $X^{n}\left(  M\right)  $ are independent, due
to the way that we selected the code (each codeword is selected independently
of a different one). The second equality follows from averaging the state
$\phi_{X^{n}}$ with respect to the distribution $p_{X^{n}}$, and we drop the
expectation $\mathbb{E}_{M}$ because the quantities inside the trace no longer
have a dependence on the message $M$. The second inequality follows from the
entropy bound for the eigenvalues of $\rho^{\otimes n}$ in the typical
subspace. The final inequality follows because Tr$\left\{  \rho^{\otimes n}%
\Pi\right\}  \leq1$.

Thus, the overall upper bound on the error probability with this sequential
decoding strategy is%
\[
\epsilon^{\prime}\equiv\epsilon+2\sqrt{\epsilon}+2\sqrt{2\sqrt{\epsilon
}+2^{-n\left[  H\left(  \rho\right)  -\delta\right]  }\ \left\vert
\mathcal{M}\right\vert },
\]
which we can make arbitrarily small by choosing $\left\vert \mathcal{M}%
\right\vert =2^{n\left[  H\left(  \rho\right)  -2\delta\right]  }$ and $n$
sufficiently large. The next arguments are standard. We proved a bound on the
expectation of the average probability, which implies there exists a
particular code that has arbitrarily small average error probability under the
same choice of $\left\vert \mathcal{M}\right\vert $ and $n$. For this code, we
can then eliminate the worst half of the codewords, ensuring that the error
probability of the resulting code is no larger than $2\epsilon^{\prime}$.
Furthermore, it should be clear that it is only necessary for the sequential
decoder to process the remaining codewords when decoding messages.
\end{proof}

\begin{remark}
Sen's proof applies to the more general case of classical-quantum channels
$x\rightarrow\rho_{x}$, with $\rho_{x}$ a mixed state, by employing
conditionally typical projectors \cite{S11}. For pure-state classical-quantum
channels, the conditionally typical projector is just the pure state itself,
and the proof simplifies as seen above.
\end{remark}

\end{document}